\documentclass{article}

\usepackage{arxiv}

\usepackage[utf8]{inputenc}
\usepackage[T1]{fontenc}
\usepackage{amsmath}
\usepackage{amssymb}
\usepackage{amsthm}
\usepackage{mathtools}
\usepackage{booktabs}
\usepackage{graphicx}
\usepackage{subcaption}
\usepackage{xcolor}
\usepackage{tcolorbox}
\usepackage{enumitem}
\usepackage{microtype}
\usepackage{url}
\usepackage{cite}
\usepackage{hyperref}
\providecommand{\headeright}{}
\renewcommand{\headeright}{}
\providecommand{\undertitle}{}
\renewcommand{\undertitle}{}
\hypersetup{
    colorlinks=true,
    linkcolor=black,
    citecolor=black,
    urlcolor=blue,
    pdftitle={Transferring Controllers from Math to Real-World Systems: A Formal Control-Theoretic Approach},
    pdfauthor={P Sangeerth, Pushpak Jagtap}
}

\graphicspath{{./images/}}

\theoremstyle{plain}
\newtheorem{theorem}{Theorem}[section]
\newtheorem{lemma}[theorem]{Lemma}

\theoremstyle{definition}
\newtheorem{definition}[theorem]{Definition}
\newtheorem{assumption}[theorem]{Assumption}
\newtheorem{problem}[theorem]{Problem}
\theoremstyle{remark}
\newtheorem{remark}[theorem]{Remark}

\title{Transferring Controllers from Math to Real-World Systems:\\ A Formal Control-Theoretic Approach}

\author{
  P Sangeerth \\
  Centre for Cyber-Physical Systems\\
  Indian Institute of Science\\
  Bengaluru, India \\
  \texttt{sangeerthp@iisc.ac.in} \\
  \And
Bhabani Shankar Dey \\
   School of Electrical Engineering, \\
   Vellore Institute of Technology Vellore,\\
  India \\
  \texttt{bhabani.shankar@vit.ac.in} \\
  \And
  Pushpak Jagtap \\
  Centre for Cyber-Physical Systems\\
  Indian Institute of Science\\
  Bengaluru, India \\
  \texttt{pushpak@iisc.ac.in} \\
}

\begin{document}
\pagestyle{empty}
\maketitle
\thispagestyle{empty}

\begin{abstract}
This paper aims to improve the math-to-real or math-to-sim transfer of a mission controller designed for a Temporal Reach-Avoid-Stay (T-RAS) specification. Our approach uses the control-theoretic concept of incremental input-to-state stability ($\delta$-ISS). We model the real-world system as an uncertain system, represented by a nominal mathematical model with an additive disturbance. Our method involves two steps: first, we render the nominal mathematical model and the real-world system $\delta$-ISS; second, we design a mission controller for the mathematical model that satisfies a modified specification. We then prove that applying the same mission-level input sequence to the real-world system guarantees that the original specification is satisfied in the real world. Videos of real-world experiments are available at \url{https://youtu.be/NvjkIZxtnPM}.
\end{abstract}


\section{Introduction}

\begin{figure}[t]
    \centering
    \includegraphics[width=0.8\linewidth]{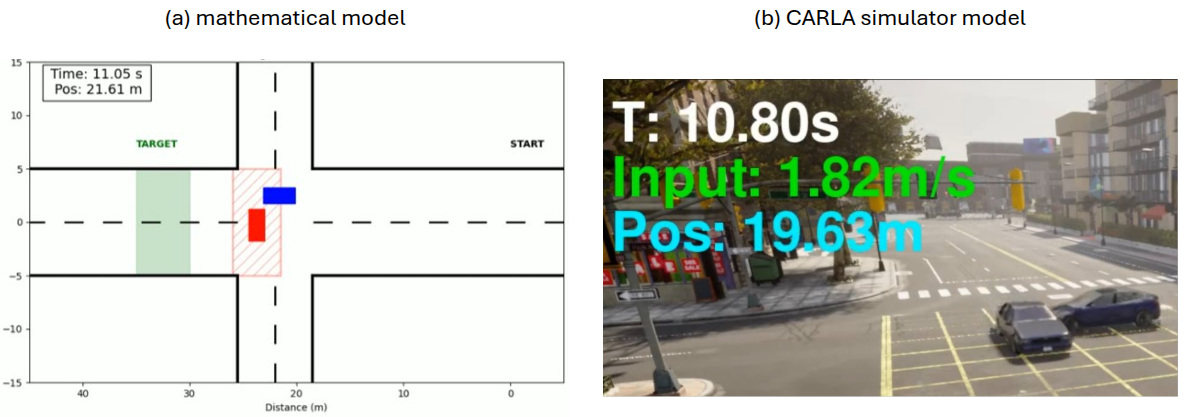}
    \caption{Mathematical controller fails when transferred. In (a), the red block represents a timed obstacle, and the blue block represents the vehicle under consideration.}
    \label{fig: carla motivate}
\end{figure}

A vast amount of literature exists on designing controllers for systems with precise mathematical models \cite{Khalil:1173048}. However, obtaining such models for real-world systems is difficult, making it challenging to transfer model-based controllers to practice or even to high-fidelity simulators. For example, a controller synthesized using a mathematical model of a car may satisfy a reach-avoid specification but fail in the CARLA simulator \cite{dosovitskiy2017carla} due to unmodeled dynamics, leading to a collision (illustrated in Fig.~\ref{fig: carla motivate}). This model-reality gap necessitates robust control-theoretic methods for enforcing mission specifications when exact models are unavailable.

To mitigate this gap, learning-based methods using reinforcement learning \cite{hasanbeig2019reinforcement} and Gaussian processes \cite{jagtap2020control,8909368,lederer2021gaussian} offer probabilistic guarantees, yet often demand specific structural properties or extensive data collection. Concurrently, broader Sim2Real transfer approaches \cite{DBLP:journals/corr/TobinFRSZA17,akella2023safety,sangeerth2024towards} employ transfer learning \cite{pmlr-v270-he25a}, bi-level optimization \cite{anand2025closing}, or deterministic gap quantification \cite{sangeerth2024towards,sangeerth2025quantification}. However, these methods generally treat the model discrepancy as additive uncertainty, forcing the design of robust controllers for uncertain systems and severely complicating the synthesis process.

A preferred alternative is to synthesize the mission controller on the deterministic mathematical model, where established formal tools provide strict guarantees, and then transfer it to the real system. Symbolic control based on simulation relations \cite{tabuada2009verification} provides such refinement guarantees under bounded disturbances through finite-state abstractions. However, the quantization of the state and input spaces leads to the curse of dimensionality and memory-intensive look-up-table controllers. Moreover, timed tasks such as Temporal Reach-Avoid-Stay (T-RAS) \cite{das2024prescribed} are typically handled by synthesizing without timing constraints and enforcing them through post-processing. Hierarchical control via approximate simulation, proposed in \cite{girard2009hierarchical}, relates a concrete system to a lower-dimensional continuous abstraction via a simulation function and an interface controller, so that controllers designed on the abstraction transfer to the concrete system with bounded error. Building on this, \cite{fainekos2009temporal} handles temporal logic specifications on a simplified model and tightens the workspace using the resulting error bound. However, both are developed for linear dynamics, consider untimed specifications, assume no mismatch between the model and the real system, and require the abstract state to be available online for the interface controller. Robust tube-based MPC \cite{LANGSON2004125} does handle bounded additive disturbances by forcing the real system to track a nominal trajectory through an ancillary controller, but it likewise relies on the nominal state online and does not explicitly address timed specifications.

To address these limitations, we propose a two-step incremental input-to-state stability ($\delta$-ISS) framework for transferring controllers from a mathematical model to a real-world system (or simulator) with formal guarantees for T-RAS specifications. First, using methods from \cite{zamani2011backstepping,basu2026neural,sangeerth2025controller,jagtap2017backstepping,zaker2025certified}, we design a feedback controller that renders both the nominal and the real-world closed-loop systems $\delta$-ISS. Second, we use the resulting trajectory-closeness bound to modify the original T-RAS specification offline and synthesize a mission controller for this modified specification directly on the nominal model. Our main contribution is a proof that applying the resulting mission-level input sequence to the real-world system guarantees satisfaction of the original timed T-RAS specification. Rather than embedding the model-reality gap in the dynamics used for synthesis, we map it into the specification. As a result, the mission controller can be synthesized with any existing method for deterministic systems. In contrast to \cite{girard2009hierarchical,fainekos2009temporal}, our framework accounts for an additive model-reality gap and preserves timed specifications; moreover, it requires no online interface controller depending on the nominal state, since the real system only needs the $\delta$-ISS feedback acting on its own state.

\section{Preliminaries}

\textbf{Notation:}
We denote the sets of nonnegative integers, positive real numbers, and nonnegative real numbers by $\mathbb{N}:=\{0,1,2,\dots\}$, $\mathbb{R}^+$, and $\mathbb{R}^+_0$, respectively. The symbol $\mathbb{R}^n$ denotes the $n$-dimensional Euclidean space. A column vector $x \in \mathbb{R}^n$ is denoted by $x=[x_1;x_2;\cdots;x_n]$. The Euclidean norm of $x \in \mathbb{R}^n$ is denoted by $\lVert x\rVert$. For any function $\phi:\mathbb{N} \rightarrow \mathbb{R}^n$, we define $\lVert \phi \rVert_{\infty}:=\sup\{\lVert \phi(k) \rVert : k \in \mathbb{N}\} \leq \infty$. We use boldface symbols (e.g., $\boldsymbol{x}$) to denote trajectories or sequences evolving over time, and plain symbols (e.g., $x \in X$) to denote a particular value $x$ in some space $X$. For $a,b \in \mathbb{N}$ with $a<b$, we define $[a,b]_{\mathbb{N}}:=\{a,a+1,\dots,b\}$. A continuous function $\alpha:\mathbb{R}^+_0\rightarrow\mathbb{R}^+_0$ is of class $\mathcal{K}$ if $\alpha(0)=0$ and it is strictly increasing. If $\alpha\in\mathcal{K}$ is radially unbounded, i.e., $\alpha(r)\rightarrow\infty$ as $r\rightarrow\infty$, then $\alpha$ is of class $\mathcal{K}_{\infty}$. A continuous function $\beta:\mathbb{R}^+_0\times\mathbb{R}^+_0\rightarrow\mathbb{R}^+_0$ is of class $\mathcal{KL}$ if, for each fixed $s$, $\beta(\cdot,s)\in\mathcal{K}$, and for each fixed $r$, $\beta(r,\cdot)$ is decreasing with $\beta(r,s)\rightarrow 0$ as $s\rightarrow\infty$. A ball centered at $x \in \mathbb{R}^n$ with radius $\epsilon \in \mathbb{R}^+$ is defined as $\mathcal{B}(x,\epsilon):=\{z\in \mathbb{R}^n \mid \lVert z-x\rVert \leq \epsilon\}$.

\subsection{System Description}
\label{system definitions subsection}
Consider a real-world system whose mathematical model is represented by a discrete-time control system (dt-CS)
\begin{equation}
    \label{eqn: math wo inc stable}
    \Sigma_m: \boldsymbol{x}(k+1)=f^m(\boldsymbol{x}(k),\boldsymbol{u}(k)),
\end{equation}
where $\boldsymbol{x}:\mathbb{N}\rightarrow X$ denotes the state of the mathematical model, $X \subset \mathbb{R}^n$ is a compact state space, $\boldsymbol{u}:\mathbb{N}\rightarrow U$ is the applied control input, $U\subset \mathbb{R}^m$ is the input set, and $f^m: X \times U \rightarrow X$ is the transition map of $\Sigma_m$.

The corresponding real-world system and/or its simulator model is represented by the nonlinear dt-CS
\begin{equation}
\label{eqn: open loop real world}
\Sigma_r: \boldsymbol{x}(k+1)=f^r(\boldsymbol{x}(k),\boldsymbol{u}(k),\boldsymbol{w}(k))=f^m(\boldsymbol{x}(k),\boldsymbol{u}(k))+\boldsymbol{w}(k),
\end{equation}
where $\boldsymbol{w}:\mathbb{N} \rightarrow W$ represents the \emph{additive} uncertainty, which captures the gap between the nominal mathematical model and the real-world system. This uncertainty does not capture measurement noise, sensing latency, bias, or state-estimation errors unless they are explicitly modeled.

We now consider the mathematical model \eqref{eqn: math wo inc stable} under a feedback controller $g:X \times \Upsilon \rightarrow U$. The resulting closed-loop system is
\begin{equation}
\label{eqn: closed loop math model}
    \Sigma_{gm}: \boldsymbol{x}(k+1) = f^m\big(\boldsymbol{x}(k), g(\boldsymbol{x}(k),\boldsymbol{\upsilon}(k))\big) = f^{gm}(\boldsymbol{x}(k),\boldsymbol{\upsilon}(k)),
\end{equation}
where $\boldsymbol{\upsilon}:\mathbb{N}\rightarrow \Upsilon$ is the new external input sequence, $\Upsilon$ is a compact external input set, and $f^{gm}:X \times \Upsilon \rightarrow X$ is the closed-loop transition map. We denote the solution of \eqref{eqn: closed loop math model} at step $k$, starting from an initial condition $x_0 \in X_0$ under an input sequence $\boldsymbol{\upsilon}$, by $\boldsymbol{x}^{gm}_{x_0 \boldsymbol{\upsilon}}(k)$, where $X_0$ is the set of initial conditions.

Similarly, applying the same controller $g$ to the real-world system \eqref{eqn: open loop real world} yields the closed-loop real-world system
\begin{align}
\label{eqn: closed loop real world}
    \Sigma_{gr}: \boldsymbol{x}(k+1) &= f^m\big(\boldsymbol{x}(k), g(\boldsymbol{x}(k),\boldsymbol{\upsilon}(k))\big) + \boldsymbol{w}(k) \nonumber \\
    &= f^{gm}(\boldsymbol{x}(k),\boldsymbol{\upsilon}(k)) + \boldsymbol{w}(k) = f^{gr}(\boldsymbol{x}(k),\boldsymbol{\upsilon}(k),\boldsymbol{w}(k)),
\end{align}
where $f^{gr}:X\times \Upsilon \times W \rightarrow X$ is the closed-loop real-world transition map. We denote the solution of \eqref{eqn: closed loop real world} at step $k$, starting from $x_0 \in X_0$ under an input sequence $\boldsymbol{\upsilon}$, by $\boldsymbol{x}^{gr}_{x_0 \boldsymbol{\upsilon}}(k)$. Next, we define the specifications intended for the closed-loop system. The primary objective of this work is to design a controller for a \emph{Temporal Reach-Avoid-Stay (T-RAS)} specification, defined as follows.

\subsection{Temporal Reach-Avoid-Stay (T-RAS) Task}

\begin{definition}[Temporal reach-avoid-stay (T-RAS) task]
    Given a state space $X$, a time instant $t_c \in \mathbb{N}$, an unsafe set $O$, an initial set $X_0 \subset X\setminus O$, a target set $T \subset X\setminus O$, and time intervals $[a,b]_{\mathbb{N}} \subseteq [0, t_c]_{\mathbb{N}}$ and $[c,d]_{\mathbb{N}} \subseteq [0, t_c]_{\mathbb{N}}$, we say that a trajectory of the system satisfies the T-RAS specification if, for all $x(0) \in X_0$, there exists $k \in [a,b]_{\mathbb{N}}$ such that $x(k) \in T$; $x(s)\notin O$ for all $s \in [c,d]_{\mathbb{N}}$; and $x(s) \in X$ for all $s \in [0,t_c]_{\mathbb{N}}$. Note that, in general, $O$ can be nonconvex and disconnected.
\end{definition}

The T-RAS task can be expressed as a metric interval temporal logic (MITL) formula \cite{karaman2008vehicle}:
\begin{equation}
\label{eqn: T-RAS Spec}
    \phi: X_0 \rightarrow \Diamond_{[a,b]_{\mathbb{N}}} T \wedge \square_{[c,d]_{\mathbb{N}}} \neg O \wedge \square_{[0,t_c]_{\mathbb{N}}} X,
\end{equation}
where $\Diamond$ and $\square$ denote the temporal operators \emph{eventually} and \emph{always} \cite{tabuada2009verification}. The task $\phi$ requires that the trajectory starting from the initial set $X_0$ eventually reaches the goal set $T \subset X$ within the interval $[a,b]_{\mathbb{N}}$, always avoids the unsafe set $O \subset X$ during $[c,d]_{\mathbb{N}}$, and always remains within the state space $X$ during $[0,t_c]_{\mathbb{N}}$. The T-RAS specification can also be expressed in signal temporal logic (STL), as STL is a subset of MITL \cite[Discussion section]{11219176}.

\subsection{Objective}

\begin{tcolorbox}[width=\linewidth]
\begin{problem}
\label{Problem-1}
Given a mathematical model $\Sigma_m$ in \eqref{eqn: math wo inc stable}, its corresponding real-world system and/or simulator model $\Sigma_r$ in \eqref{eqn: open loop real world}, and a T-RAS specification $\phi$ in \eqref{eqn: T-RAS Spec}, synthesize a controller based on the mathematical model $\Sigma_m$ that guarantees satisfaction of $\phi$ by the real-world system $\Sigma_r$.
\end{problem}
\end{tcolorbox}

During math-to-real transfer, trajectories of $\Sigma_m$ and $\Sigma_r$ inherently diverge due to unmodeled dynamics, even under identical inputs. To guarantee \emph{formal correctness transfer}, we leverage incremental input-to-state stability ($\delta$-ISS). By bounding trajectory differences as a function of initial states, input deviations, and disturbance magnitudes, $\delta$-ISS dynamically bounds the transfer gap and forms the theoretical foundation of our work.

\subsection{Incremental Input-to-State Stability}

\begin{definition}[Incremental input-to-state stability \cite{bayer2013discrete}]
\label{defn: del_iss}
The closed-loop system $\Sigma_{gr}$ in \eqref{eqn: closed loop real world} is incrementally input-to-state stable ($\delta$-ISS) if there exist functions $\beta \in \mathcal{KL}$ and $\gamma_1, \gamma_2 \in \mathcal{K}_{\infty}$ such that, for any initial conditions $x_0,y_0 \in X_0$, any input sequences $\boldsymbol{\upsilon}, \boldsymbol{\upsilon}': \mathbb{N} \rightarrow \Upsilon$, any disturbance sequences $\boldsymbol{w}, \boldsymbol{w}': \mathbb{N} \rightarrow W$, and any $k \in \mathbb{N}$,
\begin{equation}
\label{eqn: delta_isps}
\lVert \boldsymbol{x}^{gr}_{x_0\boldsymbol{\upsilon}}(k) - \boldsymbol{y}^{gr}_{y_0\boldsymbol{\upsilon}'}(k) \rVert \leq \beta(\lVert x_0-y_0\rVert, k) + \gamma_1(\lVert \boldsymbol{\upsilon} - \boldsymbol{\upsilon}' \rVert_{\infty}) + \gamma_2(\lVert \boldsymbol{w} - \boldsymbol{w}' \rVert_{\infty}).
\end{equation}
\end{definition}

Next, we introduce the $\delta$-ISS Lyapunov function for discrete-time systems, which provides sufficient conditions for the closed-loop system under controller $g$ to be $\delta$-ISS.

\begin{definition}
\label{def: delta isps Lyapunov defintion}
    For the closed-loop system $\Sigma_{gr}$ in \eqref{eqn: closed loop real world}, a continuously differentiable function $V:X \times X \rightarrow \mathbb{R}^+_{0}$ is a $\delta$-ISS Lyapunov function if there exist functions $\underline{\alpha},\bar{\alpha},\sigma,\sigma_1, \sigma_2 \in \mathcal{K}_{\infty}$ such that, for all $x,y \in X$, $\upsilon,\upsilon' \in \Upsilon$, and $w,w'\in W$,
    \begin{align}
    &\underline{\alpha}(\lVert x-y \rVert) \leq V(x,y) \leq \bar{\alpha}(\lVert x-y \rVert), \label{cdtn: first lyapunov cdtn} \\
    &V(f^{gr}(x,\upsilon,w),f^{gr}(y,\upsilon',w'))-V(x,y) \leq -\sigma(\lVert x-y \rVert) + \sigma_1(\lVert \upsilon - \upsilon' \rVert)+\sigma_2(\lVert w-w'\rVert).\label{cdtn: second lyapunov cdtn}
    \end{align}
\end{definition}

\begin{lemma}[\cite{bayer2013discrete}]
\label{lemma: delta_iss_lyapunov_sufficiency_lemma}
   If system \eqref{eqn: closed loop real world} admits a $\delta$-ISS Lyapunov function, then it is $\delta$-ISS.
\end{lemma}

Before addressing Problem~\ref{Problem-1}, we introduce a lemma that will be used in Section~\ref{section: two stage controller}.

\begin{lemma}
\label{lemma: real world delta iss}
    Consider the closed-loop systems $\Sigma_{gm}$ and $\Sigma_{gr}$ in \eqref{eqn: closed loop math model} and \eqref{eqn: closed loop real world}. If $\Sigma_{gm}$ admits a $\delta$-ISS Lyapunov function of the form $V(x,y)=\tilde{V}(x-y)$, with $\tilde{V}$ continuously differentiable, then $\Sigma_{gr}$ is $\delta$-ISS with respect to the input $\boldsymbol{\upsilon}$ and the disturbance $\boldsymbol{w}$.
\end{lemma}

\begin{proof}
Since $\Sigma_{gm}$ admits a $\delta$-ISS Lyapunov function, there exist $\underline{\alpha},\bar{\alpha},\sigma,\sigma_1 \in \mathcal{K}_{\infty}$ such that \eqref{cdtn: first lyapunov cdtn} holds and, for all $x,y \in X$ and $\upsilon_1,\upsilon_2 \in \Upsilon$,
\begin{equation}
    V(f^{gm}(x,\upsilon_1),f^{gm}(y,\upsilon_2))-V(x,y) \leq -\sigma(\lVert x-y\rVert)+\sigma_1(\lVert \upsilon_1-\upsilon_2\rVert). \label{eqn: lyap second cdtn for disturbance free sys}
\end{equation}
By hypothesis, $V(x,y)=\tilde{V}(x-y)$, which is standard in incremental stability analysis. Since $\tilde{V}$ is continuously differentiable and $X$ is compact, $\tilde{V}$ is Lipschitz on the relevant domain with some constant $L>0$. For the real-world system $\Sigma_{gr}$, consider the same function $V$. Condition \eqref{cdtn: first lyapunov cdtn} holds trivially for all $x,y \in X$. Using \eqref{eqn: lyap second cdtn for disturbance free sys}, for all $x,y \in X$, $\upsilon_1, \upsilon_2 \in \Upsilon$, and $w_1,w_2 \in W$,
\begin{align*}
    &V(f^{gr}(x,\upsilon_1,w_1),f^{gr}(y,\upsilon_2,w_2))-V(x,y)\\
    &\quad= V(f^{gm}(x,\upsilon_1)+w_1,\,f^{gm}(y,\upsilon_2)+w_2)-V(f^{gm}(x,\upsilon_1),f^{gm}(y,\upsilon_2))  + V(f^{gm}(x,\upsilon_1),f^{gm}(y,\upsilon_2))-V(x,y)\\
    &\quad\leq -\sigma(\lVert x-y\rVert)+\sigma_1(\lVert \upsilon_1-\upsilon_2\rVert)  + \tilde{V}\big(f^{gm}(x,\upsilon_1)-f^{gm}(y,\upsilon_2)+w_1-w_2\big)-\tilde{V}\big(f^{gm}(x,\upsilon_1)-f^{gm}(y,\upsilon_2)\big)\\
    &\quad\leq -\sigma(\lVert x-y\rVert)+\sigma_1(\lVert \upsilon_1-\upsilon_2\rVert)+L\lVert w_1-w_2\rVert \\
    &\quad= -\sigma(\lVert x-y\rVert)+\sigma_1(\lVert \upsilon_1-\upsilon_2\rVert)+\sigma_2(\lVert w_1 -w_2 \rVert),
\end{align*}
where $\sigma, \sigma_1 \in \mathcal{K}_{\infty}$ are as in \eqref{eqn: lyap second cdtn for disturbance free sys} and $\sigma_2(s):=Ls$. Thus, $V$ satisfies the $\delta$-ISS Lyapunov conditions for $\Sigma_{gr}$ with respect to $\boldsymbol{\upsilon}$ and $\boldsymbol{w}$, and by Lemma~\ref{lemma: delta_iss_lyapunov_sufficiency_lemma}, $\Sigma_{gr}$ is $\delta$-ISS.
\end{proof}

\section{Hierarchical Control Framework}
\label{section: two stage controller}

The mathematical model $\Sigma_m$ used for control design does not account for the real-world uncertainty $\boldsymbol{w}$, which is always present in the real-world system and/or simulator model $\Sigma_r$ \cite{sangeerth2024towards}. Consequently, controllers designed for the ideal mathematical model may fail to meet the required safety and mission goals (i.e., satisfy the T-RAS specification $\phi$) when deployed in the real world. We show that enforcing $\delta$-ISS is the key step for bridging this gap, as formally established in Theorem~\ref{theorem:specification_transfer}.

\begin{figure}[t]
    \centering
    \includegraphics[width=0.75\linewidth]{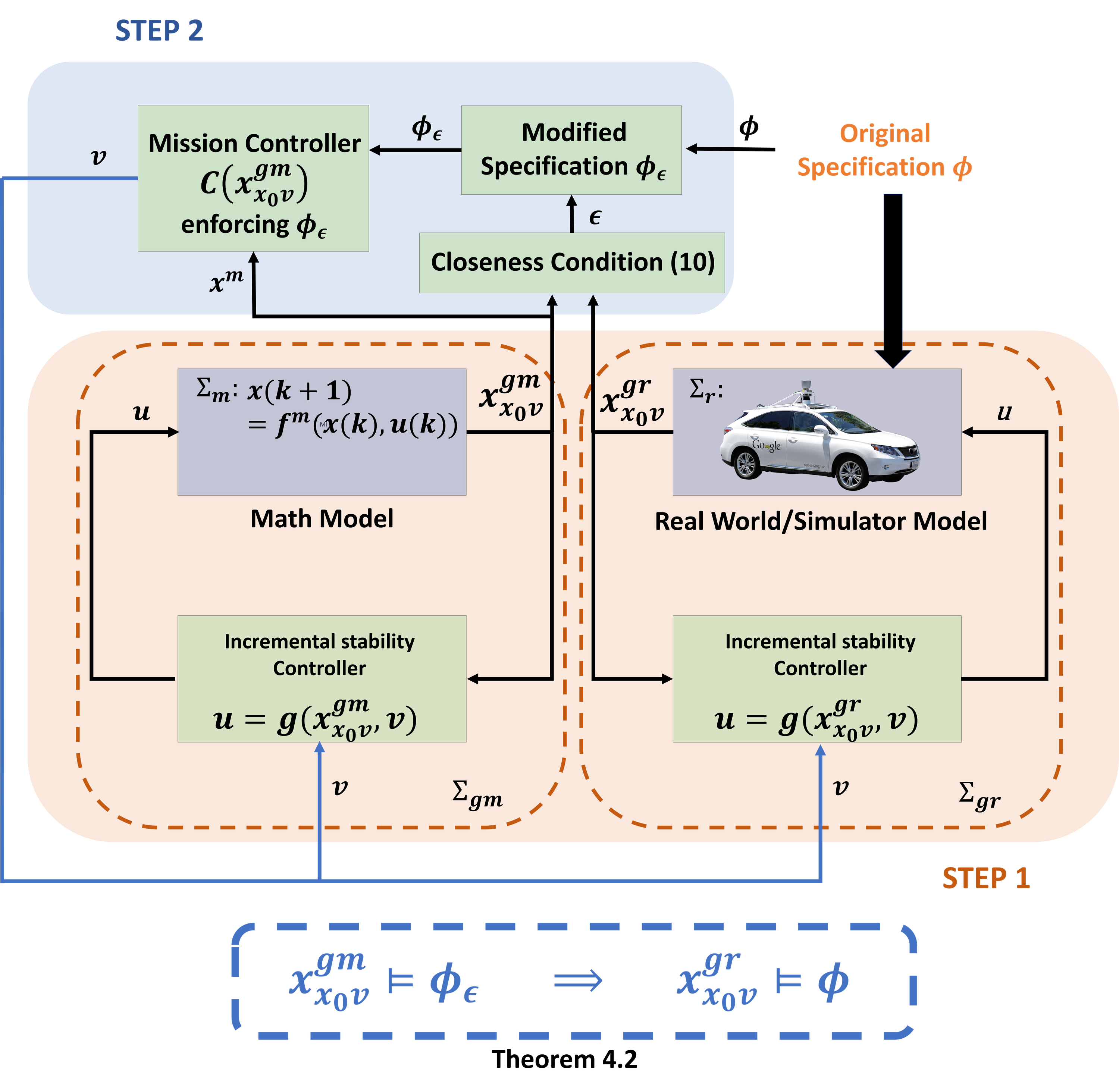}
    \caption{Two-step hierarchical controller transfer from the mathematical model to the real-world system.}
    \label{fig: deliss}
\end{figure}

Our approach follows a two-step hierarchical control design based on the $\delta$-ISS property (illustrated in Fig.~\ref{fig: deliss}):
\begin{enumerate}
    \item \textbf{Step 1 ($\delta$-ISS stabilization):} We design a controller $u:=g(x,\upsilon)$ that renders the mathematical model $\Sigma_m$ $\delta$-ISS. This controller $g$ is then transferred to the real-world system $\Sigma_r$, yielding the closed-loop systems $\Sigma_{gm}$ and $\Sigma_{gr}$, respectively. Lemma~\ref{lemma: real world delta iss} guarantees that $\Sigma_{gr}$ is $\delta$-ISS with respect to the external input $\boldsymbol{\upsilon}$ and the uncertainty $\boldsymbol{w}$.
    \item \textbf{Step 2 (Mission controller transfer):} Using the trajectory-closeness property of $\delta$-ISS, we modify the original T-RAS specification $\phi$ to a specification $\phi_{\epsilon}$ for some $\epsilon>0$ (the selection of $\epsilon$ is discussed below). A mission controller is designed for the $\delta$-ISS closed-loop mathematical model $\Sigma_{gm}$ for the modified specification $\phi_{\epsilon}$, and the resulting input sequence is applied directly to the closed-loop real-world system $\Sigma_{gr}$. The $\delta$-ISS property ensures that the real-world trajectory remains close to the trajectory of the mathematical model, thereby ensuring satisfaction of the original specification $\phi$ by the real-world system.
\end{enumerate}

Before proving how incremental stability yields closeness between the responses of $\Sigma_m$ and $\Sigma_r$, we state the following assumption.

\begin{assumption}\label{assump: all_assumption}
We make the following assumptions.
\begin{enumerate}[label=\textbf{(A\arabic*)}, ref=(A\arabic*)]
    \item There exists a full-state feedback controller $u:=g(x,\upsilon)$ that renders the mathematical model $\Sigma_m$ $\delta$-ISS with respect to $\boldsymbol{\upsilon}$. \label{assump: 1}
    \item The additive uncertainty $\boldsymbol{w}(k)\in W$, $k\in\mathbb{N}$, in $\Sigma_r$ is unknown but bounded, where $W:=\{w\in\mathbb{R}^n \mid \lVert w\rVert\le w_{max}\}$ with known $w_{max}$. \label{assump: 2}
\end{enumerate}
\end{assumption}

Assumption~\ref{assump: 1} is supported by existing $\delta$-ISS controller designs, including backstepping \cite{zamani2011backstepping} and data-driven or neural network approaches \cite{zaker2025certified,sangeerth2025controller,basu2026neural}, for broad classes of nonlinear systems. Moreover, many practical systems, such as DC motors \cite{11186986}, thermal regulation systems \cite{swikir2019compositional}, tank level systems, and traffic models \cite{sadek2023compositional}, are inherently incrementally stable. Similarly, the uncertainty bound $w_{max} := \sup_{w \in W} \lVert w \rVert$ in Assumption~\ref{assump: 2} can be estimated using existing methods such as those in \cite{alan2023control,sangeerth2024towards,sangeerth2025quantification,takano2018application}.

Since $\Sigma_{gm}$ is $\delta$-ISS with respect to $\boldsymbol{\upsilon}$, and $\Sigma_{gr}$ is $\delta$-ISS with respect to $\boldsymbol{\upsilon}$ and $\boldsymbol{w}$ (Lemma~\ref{lemma: real world delta iss}), their trajectories remain close under the same input $\boldsymbol{\upsilon}$, as quantified next.

\begin{theorem}
\label{theorem: epsilon theorem 1}
Let Assumption~\ref{assump: all_assumption} hold. Consider the $\delta$-ISS closed-loop systems $\Sigma_{gm}$ in \eqref{eqn: closed loop math model} and $\Sigma_{gr}$ in \eqref{eqn: closed loop real world}. Let $V$ be a $\delta$-ISS Lyapunov function for $\Sigma_{gr}$, characterized by class-$\mathcal{K}_{\infty}$ functions $\underline{\alpha}, \bar{\alpha}, \sigma, \sigma_1, \sigma_2$ as in Definition~\ref{def: delta isps Lyapunov defintion}, with $(\bar{\alpha}-\sigma)\in \mathcal{K}$. Let the current states $x \in X$ and $y \in X$ of $\Sigma_{gm}$ and $\Sigma_{gr}$ satisfy $\lVert x-y \rVert \leq \epsilon$, where $\epsilon$ satisfies
\begin{equation}
\label{eqn: epsilon theorem lyap condition}
\underline{\alpha}^{-1} \big( (\bar{\alpha}- \sigma)(\epsilon) + \sigma_2(w_{max}) \big) \leq \epsilon,
\end{equation}
with $w_{max}:=\sup_{w \in W} \lVert w \rVert$. Then, for all $\upsilon \in \Upsilon$ and $w \in W$,
\begin{equation*}
\lVert f^{gm}(x,\upsilon)- f^{gr}(y,\upsilon,w) \rVert \leq \epsilon.
\end{equation*}
\end{theorem}

\begin{proof}
Let $x \in X$ and $y \in X$ be the current states of $\Sigma_{gm}$ and $\Sigma_{gr}$, respectively, with $\lVert x-y \rVert \leq \epsilon$. By the $\delta$-ISS Lyapunov condition for $\Sigma_{gr}$ (Definition~\ref{def: delta isps Lyapunov defintion}),
\begin{align}
 V(f^{gr}(x, \upsilon, 0), f^{gr}(y, \upsilon,w)) &= V(f^{gm}(x, \upsilon), f^{gr}(y, \upsilon,w)) \nonumber \\
&\leq V(x, y) - \sigma(\lVert x-y \rVert) + \sigma_1(\lVert \upsilon - \upsilon \rVert) + \sigma_2(\lVert 0-w \rVert) \nonumber \\
&= V(x, y) - \sigma(\lVert x-y \rVert) + \sigma_2(\lVert w \rVert). \label{eq:proof-V-bound}
\end{align}
Using \eqref{cdtn: first lyapunov cdtn}, we obtain
\begin{align*}
\underline{\alpha}(\lVert f^{gm}(x, \upsilon)- f^{gr}(y, \upsilon,w) \rVert) &\leq V(f^{gm}(x, \upsilon), f^{gr}(y, \upsilon,w)) \\
&\leq (\bar{\alpha}- \sigma)(\lVert x-y \rVert) + \sigma_2(w_{max}) \leq (\bar{\alpha}- \sigma)(\epsilon) + \sigma_2(w_{max}).
\end{align*}
Applying $\underline{\alpha}^{-1}$ to both sides yields
\begin{equation*}
\lVert f^{gm}(x, \upsilon)- f^{gr}(y, \upsilon,w) \rVert \leq \underline{\alpha}^{-1} \big( (\bar{\alpha}- \sigma)(\epsilon) + \sigma_2(w_{max}) \big),
\end{equation*}
and condition \eqref{eqn: epsilon theorem lyap condition} gives $\lVert f^{gm}(x, \upsilon)- f^{gr}(y, \upsilon,w) \rVert \leq \epsilon$.
\end{proof}

Theorem~\ref{theorem: epsilon theorem 1} guarantees that the difference between the states of the $\delta$-ISS closed-loop mathematical model $\Sigma_{gm}$ and the real-world system $\Sigma_{gr}$ remains bounded by $\epsilon$ over a single time step, provided \eqref{eqn: epsilon theorem lyap condition} holds. This bound enables the reliable transfer of mission controllers synthesized for the mathematical model to the real system or simulator. The next section describes the specification modification required for this \emph{math-to-real} or \emph{math-to-sim} transfer while preserving satisfaction of the T-RAS specification in the real world.

\begin{remark}
    The difference-based structure of the Lyapunov function is well supported in the literature; see \cite{zamani2017towards,van2026incremental,zamani2011backstepping,sangeerth2025controller,jagtap2017backstepping} for examples. If the Lyapunov function is not of the form $V(x,y)=\tilde{V}(x-y)$, the method remains valid, but the real-world system is then incrementally input-to-state practically stable \cite{sangeerth2025controller}. In that case, the Lyapunov condition for the real-world system becomes $V(f^{gr}(x,\upsilon,w),f^{gr}(y,\upsilon',w'))-V(x,y) \leq -\sigma(\lVert x-y \rVert) + \sigma_1(\lVert \upsilon - \upsilon' \rVert)+\sigma_2\big(\sqrt{\lVert w\rVert^2 + \lVert w'\rVert^2}\big)$, where $\sigma_2(s):=Ls$ with $L$ the Lipschitz constant of $V$. Even then, the derivation in \eqref{eq:proof-V-bound} retains the same form, $V(f^{gr}(x,\upsilon,0),f^{gr}(y,\upsilon,w)) \leq V(x,y)-\sigma(\lVert x-y \rVert) + \sigma_2(\lVert w\rVert)$.
\end{remark}

\section{Transfer of Mission Controller}

Given a T-RAS specification $\phi$ in \eqref{eqn: T-RAS Spec} for the system $\Sigma_{r}$ and a margin $\epsilon > 0$ satisfying \eqref{eqn: epsilon theorem lyap condition}, we construct a modified specification $\phi_{\epsilon}$.

\subsection{Modified Specification $\phi_\epsilon$}
\label{sec: modified specification}

Consider the original specification $\phi: X_0 \rightarrow \Diamond_{[a,b]_{\mathbb{N}}} T \wedge \square_{[c,d]_{\mathbb{N}}} \neg O \wedge \square_{[0,t_c]_{\mathbb{N}}} X$, where $X_0 \subset X \setminus O$ is the initial set, $T \subset X\setminus O$ is the target set, and $O \subset X$ is the unsafe set. To construct $\phi_{\epsilon}$ for a margin $\epsilon > 0$ satisfying \eqref{eqn: epsilon theorem lyap condition}, we shrink the target set, expand the unsafe set, and shrink the state space by $\epsilon$:
\begin{align*}
    T_{\epsilon}&:=\{ x \in T \mid x' \in T \text{ for all } x' \text{ with } \lVert x-x' \rVert \leq \epsilon \},\\
    O_{\epsilon}&:=\{x \in \mathbb{R}^n \mid \exists x' \in O \text{ with } \lVert x-x' \rVert \leq \epsilon\},\\
    X_\epsilon&:=\{x \in X \mid \mathcal{B}(x,\epsilon) \subseteq X\}.
\end{align*}
The resulting modified T-RAS specification is
\begin{equation}
\label{eqn: modified specification}
    \phi_{\epsilon} : X_0 \rightarrow \Diamond_{[a,b]_\mathbb{N}} T_{\epsilon} \wedge \square_{[c,d]_\mathbb{N}} \neg O_{\epsilon}\wedge \square_{[0,t_c]_\mathbb{N}}X_\epsilon.
\end{equation}

A key requirement of the proposed framework is that the modified specification $\phi_{\epsilon}$ remains feasible for the nominal mathematical model. In practice, the conservativeness of $\phi_{\epsilon}$ increases with the size of the math-to-real gap $w_{max}$. A model that is closer to the real-world dynamics results in a smaller $w_{max}$, which yields a smaller required margin $\epsilon$ and significantly increases the likelihood that the modified specification remains feasible. Exact feasibility analysis remains an open problem due to the difficulty of satisfying spatial and temporal constraints simultaneously, and we leave it for future work.

\subsection{Controller Synthesis}

Various approaches exist for synthesizing controllers under temporal constraints. For instance, \cite{barbosa2019integrated} proposed a controller for trajectory tracking with time constraints, while \cite{das2025control} used control barrier functions (CBFs) to satisfy T-RAS specifications. For dynamic obstacle environments, \cite{9687668} proposed an online motion planning technique for MITL specifications. A control design framework for STL tasks is presented in \cite{juvvi2025signal}. Motion planning controllers for T-RAS specifications with kinodynamic constraints were developed in \cite{chatrola2025multi}, whereas \cite{saha2016milp} used mixed-integer linear programming (MILP) to design controllers for MITL specifications. Any of these methods can be employed to synthesize a mission controller $C: X \rightarrow \Upsilon$ for the closed-loop mathematical model $\Sigma_{gm}$ subject to the modified T-RAS specification $\phi_\epsilon$. The following subsection provides formal guarantees for transferring this controller to the real-world system $\Sigma_{gr}$.

\begin{remark}
Existing works \cite{sangeerth2024towards,sangeerth2025quantification,akella2023safety} incorporate the disturbance into the model and synthesize controllers for the resulting uncertain system, which complicates the design for T-RAS specifications. In contrast, our approach uses the disturbance bound only to verify \eqref{eqn: epsilon theorem lyap condition} and to modify the specification, so the mission controller is synthesized on a deterministic model.
\end{remark}

\subsection{Controller Transfer}

The main result below shows that a mission controller synthesized on the mathematical model for $\phi_\epsilon$ guarantees satisfaction of $\phi$ when transferred to the real-world system.

\begin{theorem}
\label{theorem:specification_transfer}
Let Assumption~\ref{assump: all_assumption} hold, and let $\Sigma_{gm}$ and $\Sigma_{gr}$ be the $\delta$-ISS closed-loop systems in \eqref{eqn: closed loop math model} and \eqref{eqn: closed loop real world} under the feedback controller $g$. Let $\phi$ be the T-RAS specification for $\Sigma_r$, and let $\epsilon>0$ satisfy \eqref{eqn: epsilon theorem lyap condition}. Let the mission controller $C: X \to \Upsilon$ be synthesized for $\Sigma_{gm}$ against the modified specification $\phi_\epsilon$ in \eqref{eqn: modified specification}. Fix $x_0 \in X_0$, and let $\boldsymbol{\upsilon}$ be the input sequence generated offline along the nominal trajectory, i.e., $\boldsymbol{\upsilon}(k) = C\big(\boldsymbol{x}^{gm}_{x_0\boldsymbol{\upsilon}}(k)\big)$ for all $k \in [0,t_c]_{\mathbb{N}}$. Suppose $\boldsymbol{x}^{gm}_{x_0\boldsymbol{\upsilon}} \models \phi_\epsilon$. Then, for any $\hat{x}_0 \in X_0$ with $\lVert x_0-\hat{x}_0\rVert \le \epsilon$, applying the same sequence $\boldsymbol{\upsilon}$ in open loop to $\Sigma_{gr}$ yields a trajectory $\boldsymbol{x}^{gr}_{\hat{x}_0\boldsymbol{\upsilon}}$ that satisfies $\phi$ and
\begin{equation*}
\lVert\boldsymbol{x}^{gm}_{x_0\boldsymbol{\upsilon}}(k)-\boldsymbol{x}^{gr}_{\hat{x}_0\boldsymbol{\upsilon}}(k)\rVert \le \epsilon, \quad \forall k \in [0,t_c]_{\mathbb{N}}.
\end{equation*}
\end{theorem}

\begin{proof}
Under the feedback controller $g$, the closed-loop systems $\Sigma_{gm}$ and $\Sigma_{gr}$ are $\delta$-ISS. Let the initial conditions satisfy $\lVert x_0-\hat{x}_0 \rVert \leq \epsilon$, with $\epsilon$ satisfying \eqref{eqn: epsilon theorem lyap condition}, and let $\boldsymbol{\upsilon}$ be the nominal input sequence defined in the theorem. Applying Theorem~\ref{theorem: epsilon theorem 1} inductively yields uniform trajectory closeness:
\begin{equation}\label{eq:abcd}
\lVert \boldsymbol{x}^{gm}_{x_0\boldsymbol{\upsilon}}(k)-\boldsymbol{x}^{gr}_{\hat{x}_0\boldsymbol{\upsilon}}(k) \rVert \leq \epsilon, \quad \forall k \in [0,t_c]_{\mathbb{N}}.
\end{equation}
We now verify that the real-world trajectory satisfies each component of the original T-RAS specification $\phi$.

\textbf{Avoid:} For $k\in[c,d]_\mathbb{N}$, since $\boldsymbol{x}^{gm}_{x_0\boldsymbol{\upsilon}}(k) \notin O_\epsilon$, its distance to any $\bar{x} \in O$ strictly exceeds $\epsilon$. By the reverse triangle inequality and \eqref{eq:abcd},
\begin{equation*}
\lVert \boldsymbol{x}^{gr}_{\hat{x}_0\boldsymbol{\upsilon}}(k)-\bar{x} \rVert \geq \lVert \boldsymbol{x}^{gm}_{x_0\boldsymbol{\upsilon}}(k)-\bar{x} \rVert - \lVert \boldsymbol{x}^{gm}_{x_0\boldsymbol{\upsilon}}(k)-\boldsymbol{x}^{gr}_{\hat{x}_0\boldsymbol{\upsilon}}(k) \rVert > \epsilon - \epsilon = 0.
\end{equation*}
Thus, $\boldsymbol{x}^{gr}_{\hat{x}_0\boldsymbol{\upsilon}}(k) \notin O$ for all $k\in [c,d]_\mathbb{N}$.

\textbf{Stay:} For $k\in[0,t_c]_\mathbb{N}$, the nominal trajectory satisfies $\boldsymbol{x}^{gm}_{x_0\boldsymbol{\upsilon}}(k) \in X_\epsilon$, so $\mathcal{B}(\boldsymbol{x}^{gm}_{x_0\boldsymbol{\upsilon}}(k),\epsilon) \subseteq X$. By \eqref{eq:abcd}, $\boldsymbol{x}^{gr}_{\hat{x}_0\boldsymbol{\upsilon}}(k)\in \mathcal{B}(\boldsymbol{x}^{gm}_{x_0\boldsymbol{\upsilon}}(k),\epsilon)$, and hence $\boldsymbol{x}^{gr}_{\hat{x}_0\boldsymbol{\upsilon}}(k) \in X$.

\textbf{Reach:} The nominal trajectory reaches $T_\epsilon$ at some time $k_i \in [a,b]_\mathbb{N}$, so $\mathcal{B}(\boldsymbol{x}^{gm}_{x_0\boldsymbol{\upsilon}}(k_i),\epsilon) \subseteq T$. By \eqref{eq:abcd}, $\boldsymbol{x}^{gr}_{\hat{x}_0\boldsymbol{\upsilon}}(k_i)$ lies in this ball and therefore belongs to $T$.

Hence, $\boldsymbol{x}^{gr}_{\hat{x}_0\boldsymbol{\upsilon}}$ satisfies $\phi$ and remains $\epsilon$-close to $\boldsymbol{x}^{gm}_{x_0\boldsymbol{\upsilon}}$ for all $k \in [0,t_c]_{\mathbb{N}}$.
\end{proof}

Note that the formal guarantees of Theorem~\ref{theorem:specification_transfer} hold only while $\lVert\boldsymbol{w}(k)\rVert\le w_{max}$ (Assumption~\ref{assump: 2}).

\begin{remark}
    Although the identical mission-level input sequence $\boldsymbol{\upsilon}$ is transferred to the real-world system, the resulting low-level control inputs $\boldsymbol{u}(k) = g(\boldsymbol{x}(k), \boldsymbol{\upsilon}(k))$ differ between the mathematical model and the real world, since the feedback controller $g$ is state-dependent (see Fig.~\ref{fig: deliss}).
\end{remark}

\section{Case Studies}

To demonstrate the effectiveness of the proposed framework, we present four case studies: (i) a Mecanum robot (hardware), (ii) a car model (CARLA simulator), (iii) a two-link manipulator (PyBullet simulator), and (iv) a temperature control system (hardware). All experiments were performed on a computer with an AMD Ryzen 9 5950X CPU, 128~GB RAM, and an NVIDIA RTX 3080 Ti GPU.

\begin{figure}[t]
    \centering
    \includegraphics[width=0.95\linewidth]{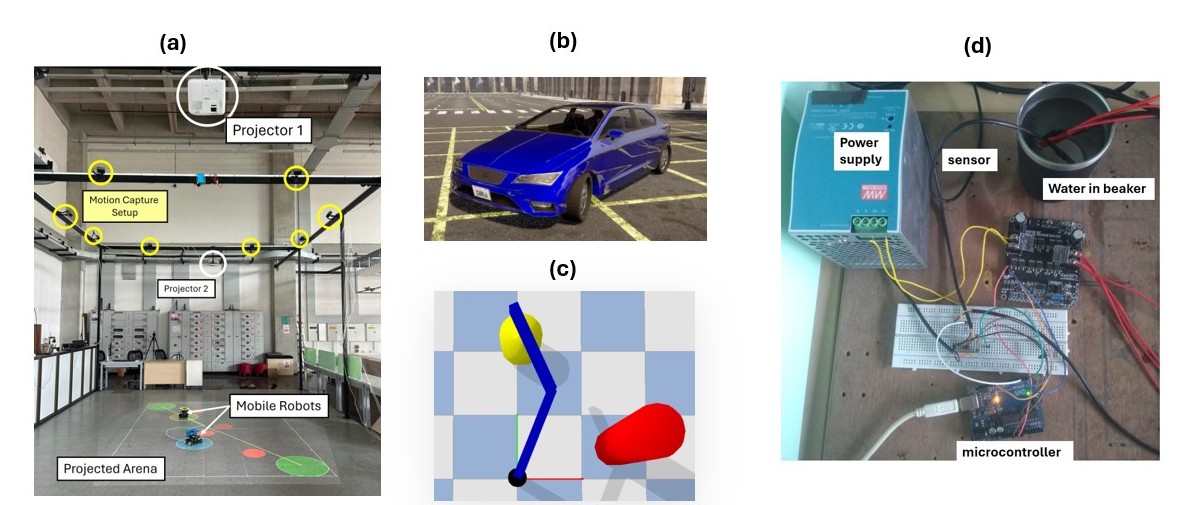}
    \caption{(a) Mecanum robot (hardware), (b) car model (CARLA simulator), (c) two-link manipulator (PyBullet simulator), and (d) temperature control system (hardware).}
    \label{fig: case study}
\end{figure}

\subsection{Mecanum Robot}

We consider a Mecanum robot with mathematical model $\Sigma_m$: $\boldsymbol{x}_1(k+1) = \boldsymbol{x}_1(k) + \tau \boldsymbol{u}_1(k)$, $\boldsymbol{x}_2(k+1) = \boldsymbol{x}_2(k) + \tau \boldsymbol{u}_2(k)$, where $\boldsymbol{x}_1, \boldsymbol{x}_2$ denote the $x$ and $y$ positions of the robot and $\boldsymbol{u}_1, \boldsymbol{u}_2$ denote the linear velocities in the $x$ and $y$ directions. The hardware setup is shown in Fig.~\ref{fig: case study}(a). Position data of the robot are collected from a PhaseSpace motion capture system\footnote{\url{https://www.phasespace.com/}} using ROS~2. The hardware is represented by the mathematical model $\Sigma_m$ subject to additive uncertainty, with sampling period $\tau=0.1$\,s. The T-RAS specification for the real system is $\Diamond_{[0,40]} T \wedge \square_{[0,40]} X \wedge \square_{[0,40]} \neg O$.\footnote{While the formal definitions in this paper use discrete time intervals $[a,b]_{\mathbb{N}}$, the specifications in this section are stated in physical time $[a\tau,b\tau]$ (time $t = k\tau$ in seconds) for intuitive interpretation.}

\begin{figure}[t]
    \centering
    \begin{subfigure}[b]{0.48\linewidth}
        \centering
        \includegraphics[width=\linewidth]{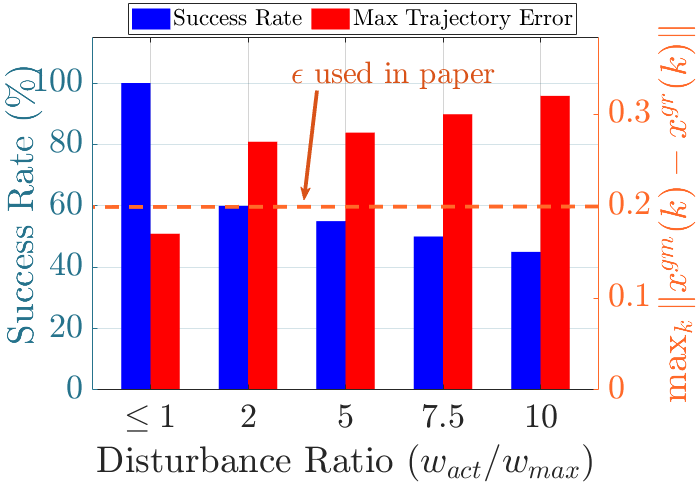}
        \caption{}
        \label{fig: psg_sensitivity_analysis_i}
    \end{subfigure}
    \hfill
    \begin{subfigure}[b]{0.48\linewidth}
        \centering
        \includegraphics[width=\linewidth]{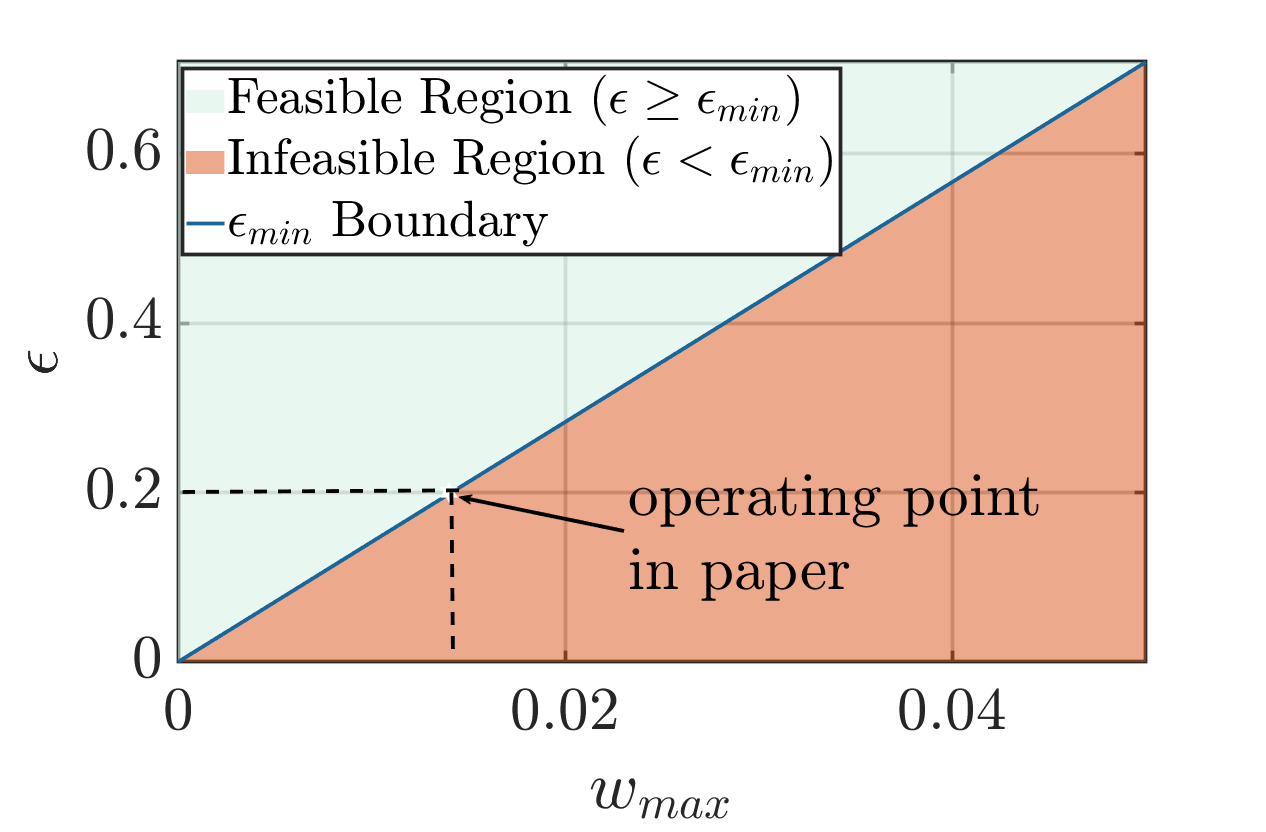}
        \caption{}
        \label{fig: psg_sensitivity_analysis_ii}
    \end{subfigure}
    \caption{Mecanum robot: (a) success rate and maximum deviation vs.\ $w_{act}/w_{max}$ (20 runs per level); (b) $\epsilon_{\min}$ vs.\ $w_{max}$ from \eqref{eqn: epsilon theorem lyap condition}.}
    \label{fig: sensitivity_analysis}
\end{figure}

\begin{figure}[t]
    \centering
    \includegraphics[width=0.95\linewidth]{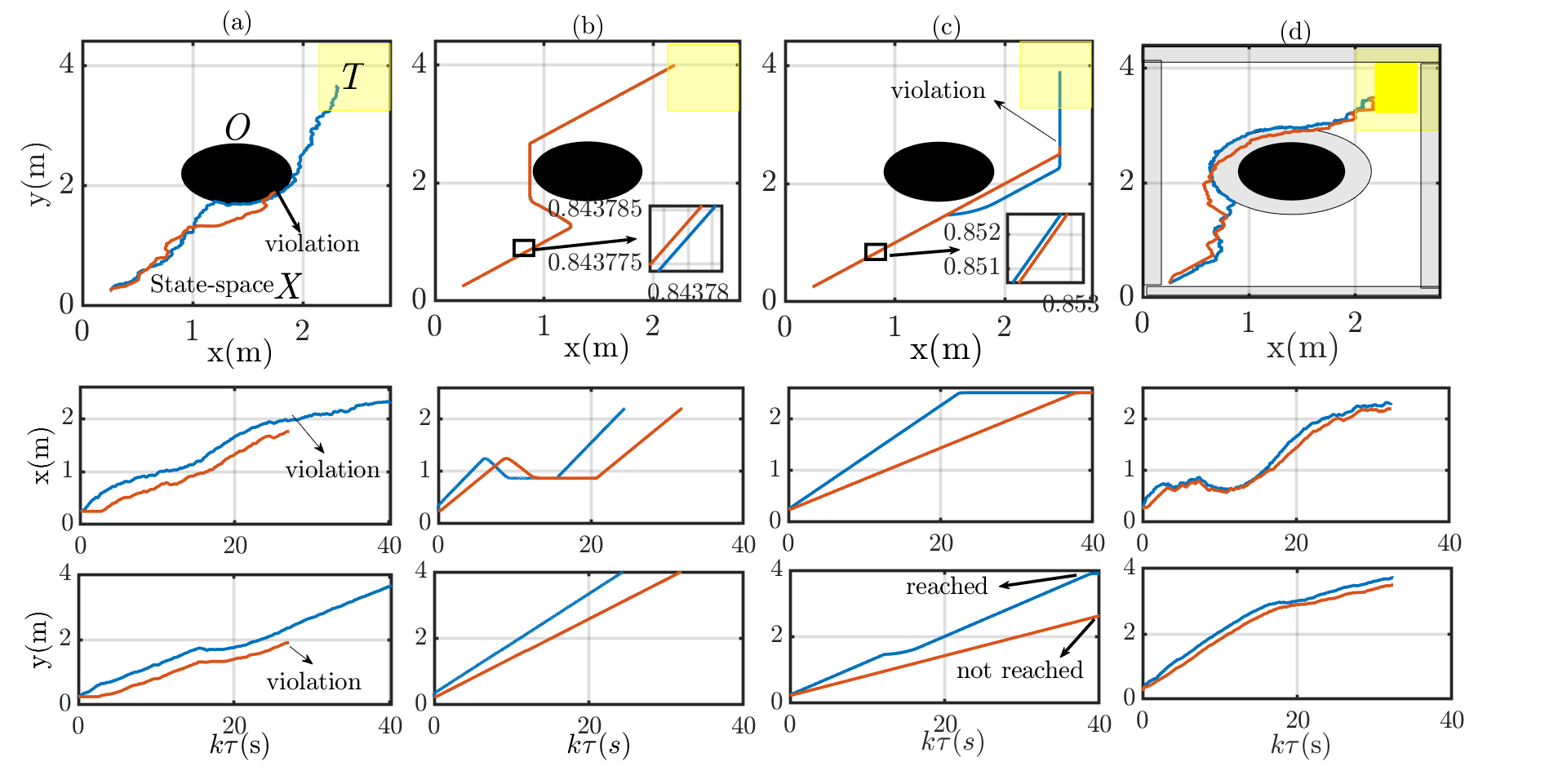}
    \caption{Trajectories of the mathematical model (blue) and hardware (red). The black region denotes the obstacle, and the yellow region denotes the target. (a) Direct transfer without $\delta$-ISS. (b) Transfer using an abstraction-based technique. (c) Transfer using tube MPC. (d) Our approach.}
    \label{fig: mecanum hardware plots}
\end{figure}

First, we synthesized a mission controller for the mathematical model using the MILP method \cite{saha2016milp} for a time horizon of $40$\,s. When deployed on the hardware, this controller failed to satisfy the specification, as shown in Fig.~\ref{fig: mecanum hardware plots}(a). To address this violation, we employed the controller $\boldsymbol{u}(k)=-\boldsymbol{x}(k)+\boldsymbol{\upsilon}(k)$ to render the mathematical model $\delta$-ISS with respect to the input $\boldsymbol{\upsilon}$. The resulting closed-loop system is characterized by the $\mathcal{K}_{\infty}$ functions $\underline{\alpha}(s)=s^2$, $\bar{\alpha}(s)=s^2$, $\sigma(s)=0.095s^2$, and $\sigma_2(s)=19.05s^2$. We empirically estimated $w_{max}=0.014$ using Gaussian process regression \cite{seeger2004gaussian}; hence, the guarantee of Theorem~\ref{theorem:specification_transfer} for this case study is conditional on this empirical estimate. Fig.~\ref{fig: psg_sensitivity_analysis_i} shows the effect of exceeding it. Deterministic GP bounds as in \cite{hashimoto2022learning} could also be used to estimate $w_{max}$, but we avoid them due to their conservatism. We obtain $\epsilon=0.2$, which satisfies \eqref{eqn: epsilon theorem lyap condition} (see Fig.~\ref{fig: sensitivity_analysis}). We then synthesized a new mission controller for the closed-loop mathematical model using the modified specification $\phi_\epsilon$ (with bloated obstacles and shrunk target and state sets) via kinodynamic RRT \cite{barbosa2019integrated}. The same input sequence was applied to the hardware system. As guaranteed by Theorem~\ref{theorem:specification_transfer}, the real-world trajectory avoided collisions and satisfied the temporal constraints (see Fig.~\ref{fig: mecanum hardware plots}(d) and the video), thus bridging the \emph{math-to-real} gap.

\textbf{Discussion:}
We benchmark the proposed framework against two existing disturbance-handling methods: the abstraction-based approach \cite{rungger2016scots} and tube-based MPC \cite{LANGSON2004125}. Although these methods are not directly designed for general T-RAS specifications \eqref{eqn: T-RAS Spec}, all temporal intervals in this case study coincide with $[0,t_c]_{\mathbb{N}}$, which allows their adaptation for comparison. Fig.~\ref{fig: mecanum hardware plots}(b) shows that the abstraction-based controller fulfills the task, but with a temporal discrepancy between the nominal and physical trajectories due to state-space discretization.\footnote{The abstraction tool \cite{rungger2016scots} does not natively support MITL specifications. However, since all temporal intervals in this case study coincide with $[0,t_c]_{\mathbb{N}}$, one can synthesize a controller for the untimed LTL counterpart of the T-RAS task and then enforce the time bound $t_c$ by post-processing the discrete transitions from the given initial state. Since this differs from direct timed synthesis, it may contribute to the observed loss of temporal robustness.} Moreover, abstraction-based methods scale poorly with the state dimension. Fig.~\ref{fig: mecanum hardware plots}(c) shows that although tube MPC satisfies the specification for the nominal model, the hardware trajectory deviates noticeably from the nominal one and violates the timing requirement, as tube MPC does not explicitly address timed specifications. In contrast, our framework formally guarantees that the real-world trajectory remains $\epsilon$-close to the nominal trajectory.

To assess sensitivity, we injected additional bounded disturbances into the input and measured the controller transfer success rate over 20 runs per disturbance level. As shown in Fig.~\ref{fig: psg_sensitivity_analysis_i}, once $w_{act}>w_{max}$, the success rate drops and the maximum deviation exceeds $\epsilon$. Fig.~\ref{fig: psg_sensitivity_analysis_ii} shows the boundary of admissible margins $\epsilon_{\min}$, i.e., the smallest $\epsilon$ satisfying \eqref{eqn: epsilon theorem lyap condition}. Table~\ref{tab: comparison_table} compares the methods in terms of computation time and spatial \cite{donze2010robust} and temporal \cite{9903850} robustness, which quantify the distance to violation in space and time, respectively. With our approach, spatial robustness decreases by 0.07 during transfer, within the $\epsilon$-closeness bound, while temporal robustness is preserved. The abstraction-based method preserves spatial robustness but loses temporal robustness, and tube MPC loses both.

\begin{table}[t]
    \centering
    \caption{Robustness and computation time comparison.}
    \label{tab: comparison_table}
    \resizebox{\linewidth}{!}{%
    \begin{tabular}{lcccccc}
        \toprule
        & \multicolumn{2}{c}{\textbf{Spatial Robustness \cite{donze2010robust}}} & \multicolumn{2}{c}{\textbf{Temporal Robustness \cite{9903850}}} & \multicolumn{2}{c}{\textbf{Computation Time}} \\
        \cmidrule(r){2-3} \cmidrule(lr){4-5} \cmidrule(l){6-7}
        \textbf{Method} & \textbf{Math} & \textbf{Real} & \textbf{Math} & \textbf{Real} & \textbf{Online} & \textbf{Offline} \\
        \midrule
        Direct transfer & 0.05 & $-0.01$ & 0 & -- & 40\,$\mu$s & 0.5\,s \\
        Abstraction-based \cite{rungger2016scots} & 0.10 & 0.10 & 17 & 8 & 60\,$\mu$s & 1\,min \\
        Tube MPC \cite{LANGSON2004125} & 0.25 & 0.05 & 2 & 0 & 0.1\,ms & -- \\
        \textbf{Our approach} & \textbf{0.27} & \textbf{0.20} & \textbf{9} & \textbf{9} & \textbf{50\,$\mu$s} & \textbf{0.56\,s} \\
        \bottomrule
    \end{tabular}%
    }
\end{table}

\subsection{CARLA Simulator Model}

\begin{figure}[t]
    \centering
    \includegraphics[width=0.75\linewidth]{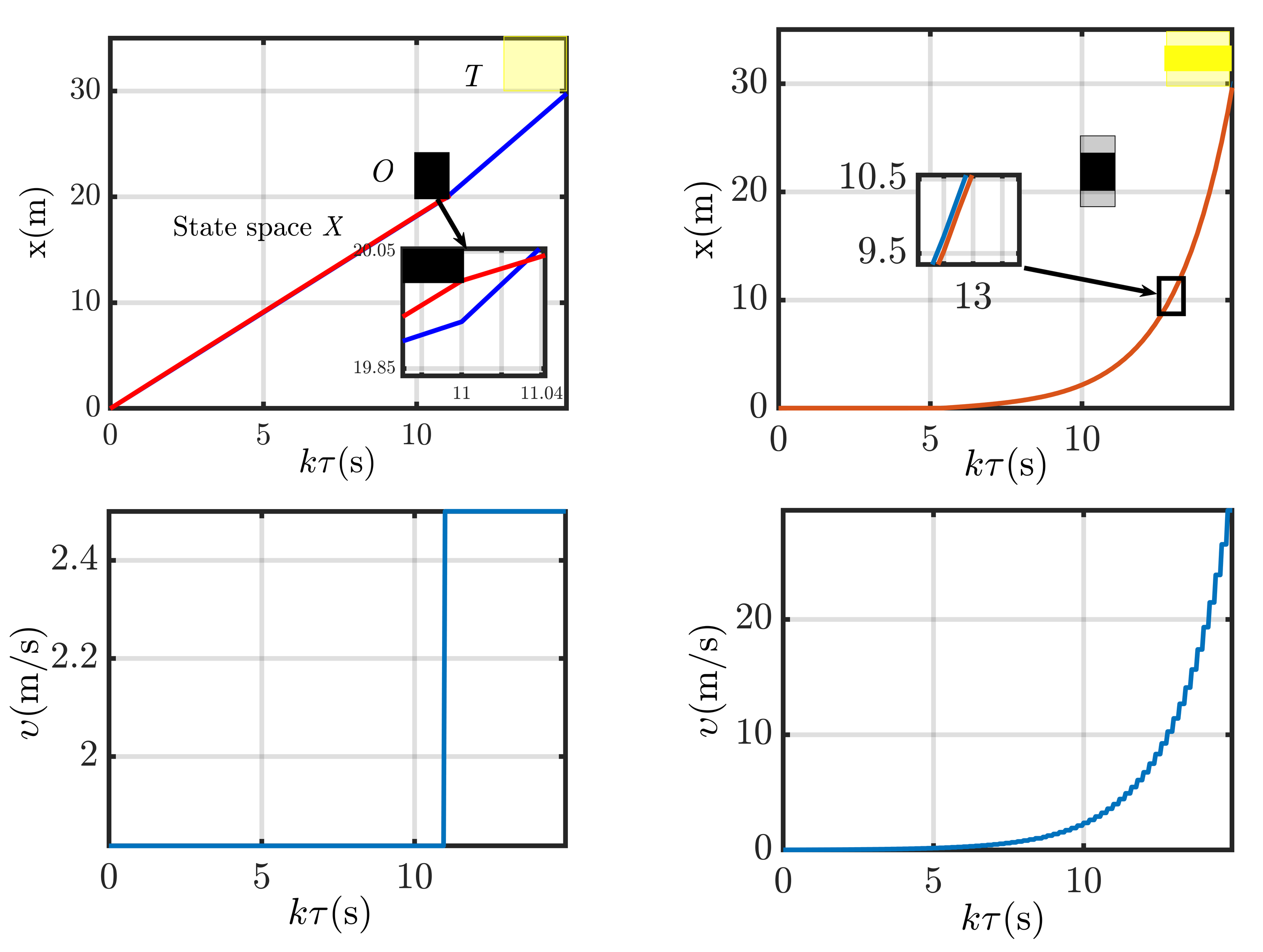}
    \caption{Mathematical model (blue) and CARLA simulator (red) state responses (top). In the top right, the obstacle is bloated and the target set is shrunk. The bottom plots show the input applied to the CARLA model without (left) and with (right) our approach.}
    \label{fig: carla input comparison}
\end{figure}

This case study demonstrates controller transfer in a safety-critical traffic scenario involving a main vehicle and an obstacle vehicle at a $90^\circ$ intersection. The kinematic single-track model from \cite{szecsi2024deep}, linearized around the $x$-direction, is used for the vehicle dynamics (see Fig.~\ref{fig: carla motivate}). The system is simulated in CARLA \cite{dosovitskiy2017carla}, as shown in Fig.~\ref{fig: case study}(b), with motion restricted to one dimension. The T-RAS specification is $\Diamond_{[13,15]} T \wedge \square_{[0,15]} X \wedge \square_{[10,11]} \neg O$. Initially, we synthesized a mission controller for the nominal mathematical model using MILP \cite{saha2016milp} for a time horizon of $15$\,s. When deployed in the simulator, this baseline controller failed to avoid the obstacle, as shown in Fig.~\ref{fig: carla input comparison} (top left). To address this, we employed a feedback controller similar to that of the previous case study to render the mathematical model $\delta$-ISS with respect to $\boldsymbol{\upsilon}$. The disturbance bound was estimated as $w_{max}=0.05$ using the method in \cite{sangeerth2024towards}, and we obtain $\epsilon=0.99$, which satisfies \eqref{eqn: epsilon theorem lyap condition}. We then synthesized a new mission controller for the closed-loop mathematical model against $\phi_\epsilon$ using MILP. As guaranteed by Theorem~\ref{theorem:specification_transfer}, the simulated trajectory avoided collisions and satisfied the temporal constraints (Fig.~\ref{fig: carla input comparison}, top right), bridging the \emph{math-to-sim} gap. Fig.~\ref{fig: carla input comparison} (bottom) compares the control efforts: our approach maintains a conservative input profile during the obstacle-avoidance phase and increases control authority only once the obstacle is cleared to meet the timing requirement, whereas the baseline uses a sharp input spike to meet the deadline, causing a collision.

\subsection{Two-Link Manipulator}

\begin{figure}[t]
    \centering
    \includegraphics[width=0.6\linewidth]{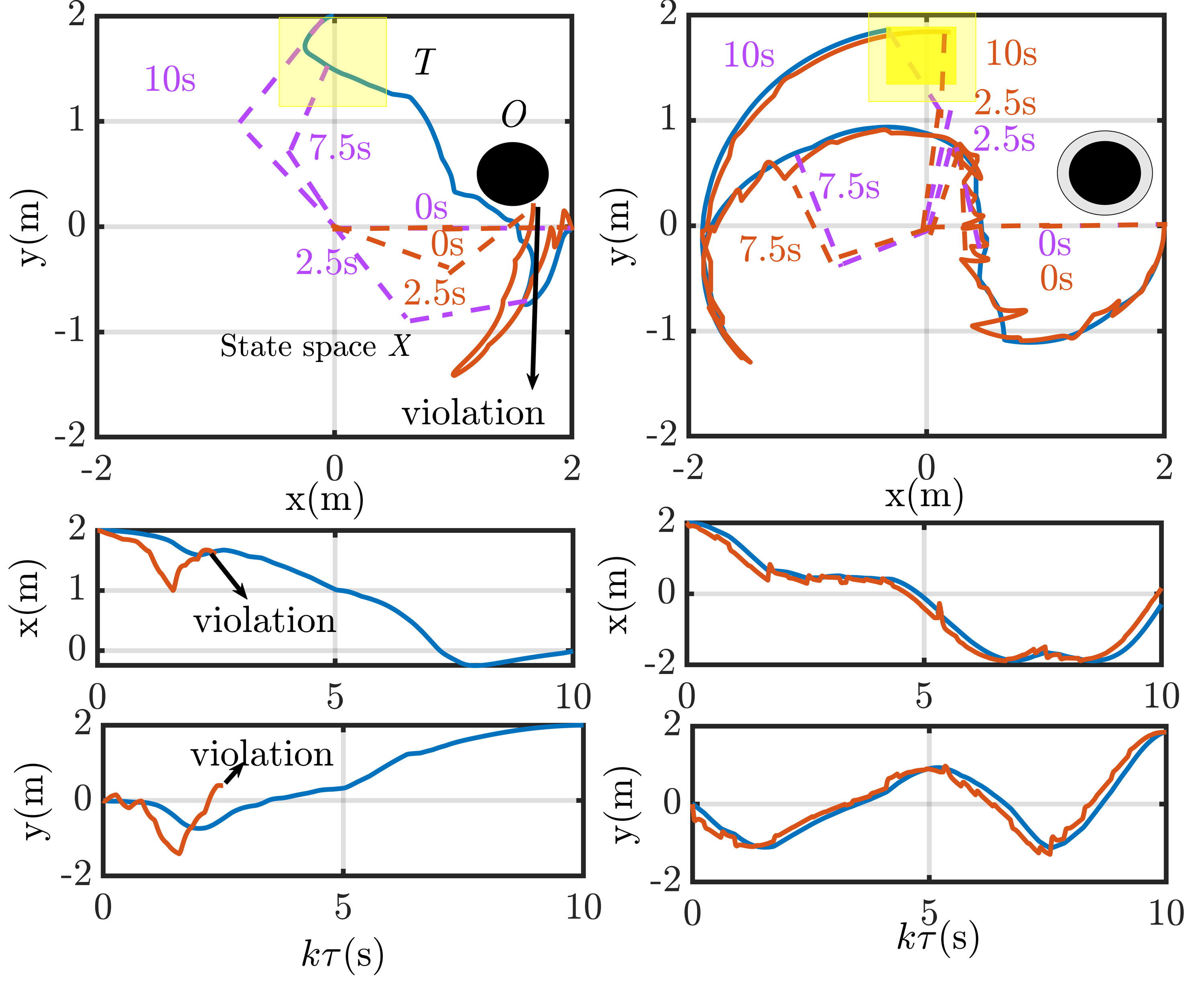}
    \caption{Two-link manipulator: model (blue) vs.\ PyBullet (red). Left: direct transfer collides. Right: the proposed approach satisfies $\phi$. Top: task-space paths; bottom: joint positions vs.\ time.}
    \label{fig: 2dof manipulator}
\end{figure}

We consider a two-link manipulator \cite{2Regbook} described by $\dot \xi_1=\xi_2$, $\dot \xi_2 = M^{-1}(\xi_1)\left[-H(\xi_1,\xi_2)-c(\xi_1)\right]+M^{-1}(\xi_1)\tau$, where $\xi=[\xi_1^{\top},\xi_2^{\top}]^{\top}$, $\xi_1=\mathbf{p}$, $\xi_2=\dot{\mathbf{p}}$, $\mathbf{p}(t)=[x_1,x_2]^{\top}$, $x_1$ and $x_2$ are the angles of the two revolute joints, and $\tau$ denotes the joint torque inputs. The parameter matrices are chosen as in \cite{2Regbook}, with link masses $m=1$\,kg and lengths $l=1$\,m. The corresponding simulator model is implemented in PyBullet \cite{coumans2021}, as shown in Fig.~\ref{fig: case study}(c). The specification is $\Diamond_{[0,10]} T \wedge \square_{[0,10]} X \wedge \square_{[0,10]} \neg O$. Fig.~\ref{fig: 2dof manipulator} compares two cases. Without our approach, when the controller is synthesized using RRT* \cite{karaman2011sampling} for a prescribed time of $10$\,s, the manipulator hits the obstacle. We use the method in \cite{sangeerth2025controller} to render the system $\delta$-ISS and obtain the functions $\underline{\alpha},\bar{\alpha},\sigma,\sigma_2$ as in the previous case studies. We compute $w_{max}=0.13$ using \cite{sangeerth2024towards} and $\epsilon=0.42$, which satisfies \eqref{eqn: epsilon theorem lyap condition}. With our approach, the mission input sequence designed for the mathematical model satisfies the T-RAS specification in the simulator, thereby bridging the \emph{math-to-sim} gap.

\subsection{Temperature Control}

\begin{figure}[t]
    \centering
    \includegraphics[width=0.6\linewidth]{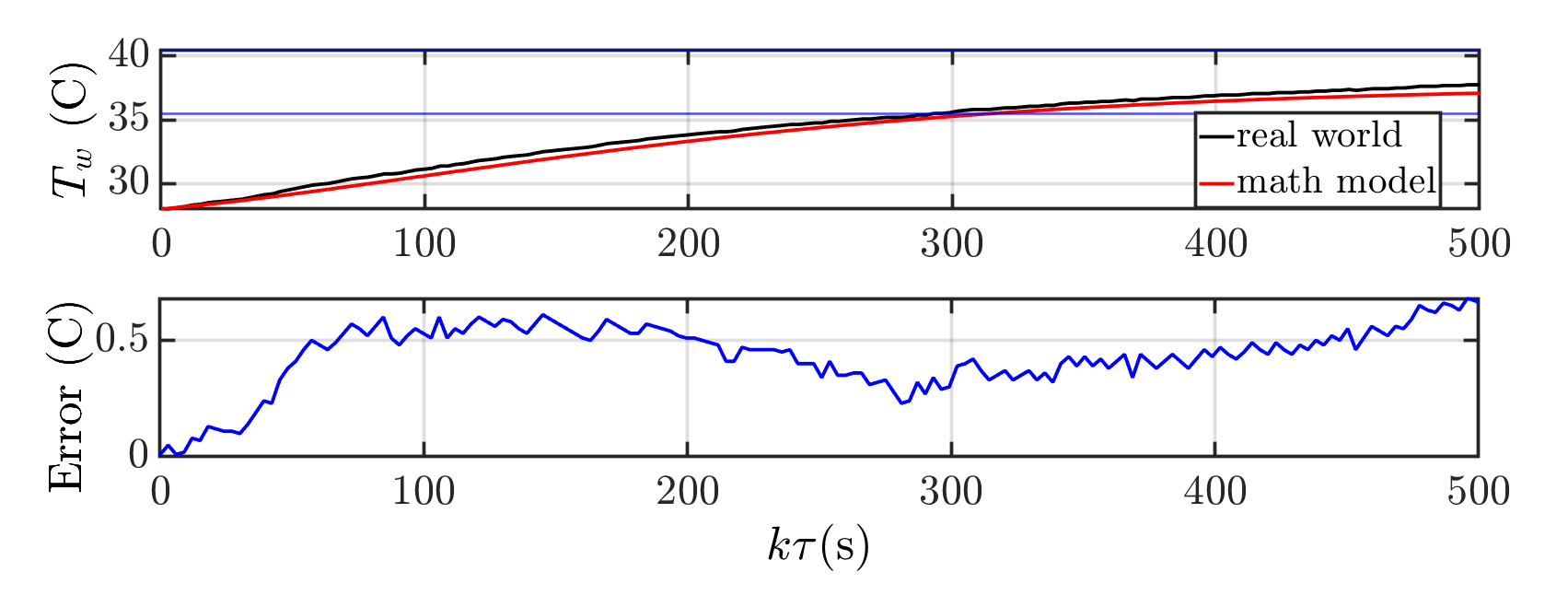}
    \caption{Temperature control system satisfying $\Diamond_{[0,500]} T$.}
    \label{fig: temperature control plots}
\end{figure}

We consider a real-time Arduino-based temperature control system (Fig.~\ref{fig: case study}(d)) governed by
\begin{align*}
\boldsymbol{T_w}(k+1)&=\boldsymbol{T_w}(k)+\Delta t\big(-a(\boldsymbol{T_w}(k)-T_{amb})-b(\boldsymbol{T_h}(k)-\boldsymbol{T_w}(k))\big),\\
\boldsymbol{T_h}(k+1)&=\boldsymbol{T_h}(k)+\Delta t \big(b(\boldsymbol{T_h}(k)-\boldsymbol{T_w}(k))+c\, \boldsymbol{Q_{in}}(k)\big),
\end{align*}
where $a,b,c>0$ are system parameters, and $\boldsymbol{T_w}$, $\boldsymbol{T_h}$, $\boldsymbol{Q_{in}}$, $T_{amb}$, and $\Delta t$ denote the water temperature, heater temperature, applied heat input, ambient temperature, and sampling period, respectively. This system is inherently $\delta$-ISS. As in the previous case studies, we compute $w_{max}=0.75$, identify the class-$\mathcal{K}_{\infty}$ functions associated with a Lyapunov function, and find that $\epsilon=2$ satisfies \eqref{eqn: epsilon theorem lyap condition}. The specification is $\Diamond_{[0,500]} T$, where $T= [36,40]\times [0,100]$. By Theorem~\ref{theorem:specification_transfer}, applying the input sequence designed for the mathematical model to the real system satisfies the reachability specification within $500$\,s, as confirmed in Fig.~\ref{fig: temperature control plots}.

\section{Conclusion and Future Work}
\label{sec: conclusion}

We presented a $\delta$-ISS-based control framework that bridges the \emph{math-to-real} and \emph{math-to-sim} gaps with formal guarantees. By rendering both the mathematical model and the real-world system $\delta$-ISS, we ensure trajectory closeness, which allows a controller synthesized on the mathematical model for a modified T-RAS specification to be transferred directly to the real world. Experiments across several case studies demonstrate its practical efficacy compared with existing methods. In future work, we aim to handle stochastic disturbances and output-feedback control.

\bibliographystyle{ieeetr}
\bibliography{sources}

\end{document}